\documentclass[aps,prc,twocolumn,showpacs,nofootinbib,superscriptaddress]{revtex4} 
\usepackage{graphicx,longtable} 
\usepackage{dcolumn}
\usepackage{mathptmx}
\usepackage{amsmath}
\usepackage[pdftex]{hyperref}
\usepackage{wasysym}
\usepackage[sort&compress]{natbib}
\usepackage{longtable,dcolumn,epsfig,color}
\usepackage{braket}
\usepackage{natbib}

\newcommand{\beqy}{\begin{eqnarray}}
\newcommand{\eeqy}{\end{eqnarray}}
\newcommand{\bmlet}{\begin{subequations}}
\newcommand{\emlet}{\end{subequations}}
\newcounter{saveeqn}

\def\gsimeq{\,\,\raise0.14em\hbox{$>$}\kern-0.76em\lower0.28em\hbox  
{$\sim$}\,\,}  
\def\lsimeq{\,\,\raise0.14em\hbox{$<$}\kern-0.76em\lower0.28em\hbox  
{$\sim$}\,\,}

\def\gsimeq{\,\,\raise0.14em\hbox{$>$}\kern-0.76em\lower0.28em\hbox {$\sim$}\,\,}
\def\lsimeq{\,\,\raise0.14em\hbox{$<$}\kern-0.76em\lower0.28em\hbox {$\sim$}\,\,}

\begin{document}

\preprint{APS/123-QED}


\title{Low-energy enhancement in the $\gamma$-ray strength functions of $^{73,74}\rm{Ge}$}

\author{T.~Renstr{\o}m}
\email{therese.renstrom@fys.uio.no}
\affiliation{Department of Physics, University of Oslo, N-0316 Oslo, Norway}
\author{H.-~T.~Nyhus}
\affiliation{Department of Physics, University of Oslo, N-0316 Oslo, Norway}
\author{H.~Utsunomiya}
\affiliation{Department of Physics, Konan University, Okamoto 8-9-1, Higashinada, Kobe 658-8501, Japan}
\author{R.~Schwengner}
\affiliation{Institute of Radiation Physics, Helmholtz-Zentrum Dresden-Rossendorf, 01328 Dresden, Germany}
\author{S.~Goriely} 
\affiliation{Institut d'Astronomie et d'Astrophysique, Universit\'{e} Libre de
Bruxelles, Campus de la Plaine, CP-226, 1050 Brussels, Belgium}
\author{A.~C.~Larsen}
\affiliation{Department of Physics, University of Oslo, N-0316 Oslo, Norway}
\author{D.~M.~Filipescu}
\affiliation{ELI-NP, "Horia Hulubei" National Institute for Physics and Nuclear Engineering (IFIN-HH), 30 Reactorului, 077125 Bucharest-Magurele, Romania}
\affiliation{"Horia Hulubei" National Institute for Physics and Nuclear Engineering (IFIN-HH), 30 Reactorului, 077125 Bucharest-Magurele, Romania}
\author{I.~Gheorghe}
\affiliation{ELI-NP, "Horia Hulubei" National Institute for Physics and Nuclear Engineering (IFIN-HH), 30 Reactorului, 077125 Bucharest-Magurele, Romania}
\author{L.~A.~Bernstein}
\affiliation{Lawrence Livermore National Laboratory, 7000 East Avenue, Livermore, California 94550-9234, US}
\author{D.~L.~Bleuel}
\affiliation{Lawrence Livermore National Laboratory, 7000 East Avenue, Livermore, California 94550-9234, US}
\author{T.~Glodariu}
\affiliation{National Institute for Physics and Nuclear Engineering Horia Hulubei, str Atomistilor nr. 407, Bucharest-Magurele, P.O.BOX MG6, Romania}
\author{A.~G\"{o}rgen}
\affiliation{Department of Physics, University of Oslo, N-0316 Oslo, Norway}
\author{M.~Guttormsen}
\affiliation{Department of Physics, University of Oslo, N-0316 Oslo, Norway}
\author{T.~W.~Hagen}
\affiliation{Department of Physics, University of Oslo, N-0316 Oslo, Norway}
\author{B.~V.~ Kheswa}
\affiliation{iThemba LABS, P.O. Box 722, 7129 Somerset West, South Africa}
\author{Y.~-W.~Lui}
\affiliation{Cyclotron Institute, Texas A\&M University, College Station, Texas 77843, USA}
\author{D.~Negi}
\affiliation{iThemba LABS, P.O. Box 722, 7129 Somerset West, South Africa}
\affiliation{Centre for Excellence in Basic Sciences, Vidyanagari Campus, Mumbai 400098, India}
\author{I.~E.~Ruud}
\affiliation{Department of Physics, University of Oslo, N-0316 Oslo, Norway}
\author{T.~Shima}
\affiliation{Research Center for Nuclear Physics, Osaka University, Suita, Osaka 567-0047, Japan}
\author{S.~Siem}
\affiliation{Department of Physics, University of Oslo, N-0316 Oslo, Norway}
\author{K.~Takahisa}
\affiliation{Research Center for Nuclear Physics, Osaka University, Suita, Osaka 567-0047, Japan}
\author{O.~Tesileanu}
\affiliation{Extreme Light Infrastructure Nuclear Physics, str Atomistilor nr. 407, Bucharest-Magurele, P.O.BOX MG6, Romania}
\author{T.~G.~Tornyi}
\affiliation{Department of Physics, University of Oslo, N-0316 Oslo, Norway}
\affiliation{Institute of Nuclear Research of the Hungarian Academy of Sciences (MTA Atomki), Debrecen, Hungary}
\author{G.~M.~Tveten}
\affiliation{Department of Physics, University of Oslo, N-0316 Oslo, Norway}
\author{ M.~Wiedeking}
\affiliation{iThemba LABS, P.O. Box 722, 7129 Somerset West, South Africa}

\date{\today}


\begin{abstract}
\vskip 0.5cm
The $\gamma$-ray strength functions and level densities of $^{73,74}$Ge have been extracted up to the neutron separation energy S$_n$ from particle-$\gamma$ coincidence data using the Oslo method. Moreover, the $\gamma$-ray strength  function of $^{74}$Ge above S$_n$ has been determined from photo-neutron measurements; hence these two experiments cover the range of E$_\gamma \approx$ 1-13 MeV for $^{74}$Ge. The obtained data show that both $^{73,74}$Ge display an increase in strength at low $\gamma$ energies. The experimental $\gamma$-ray strength functions are compared with $M1$ strength functions deduced from average $B(M1)$ values calculated within the shell model for a large number of transitions. The observed low-energy enhancements in $^{73,74}$Ge are adopted in the calculations of the $^{72,73}$Ge(n,$\gamma$) cross sections, where there are no direct experimental data. Calculated reaction rates for more neutron-rich germanium isotopes are shown to be strongly dependent on the presence of the low-energy enhancement.   
\end{abstract}



\maketitle



\section{Introduction}
\label{sec:int}
A good knowledge on how the atomic nucleus emits and absorbs photons is essential for the fundamental understanding of this many-faceted quantum system, as well as for a wide range of nuclear applications. To characterize the average, nuclear response to electromagnetic radiation, the $\gamma$-ray strength function ($\gamma$SF)~\cite{Bartholomew} has proven to be a fruitful concept when the nucleus is excited to high energies, and the density of quantum levels is high. There exists a wealth of information about the $\gamma$SF for nuclei above the neutron binding energy, S$_n$, predominantly from photo-neutron experiments~\cite{atlasD&B} and from the spectrum-fitting method~\cite{SchillerT}. For $\gamma$ energies below $S_n$ the information is more scarce, as it remains quite challenging to extract the $\gamma$SF experimentally in this energy range. For this region the Oslo method~\cite{schiller_geni}, the two-step cascade method~\cite{Hoogenboom} and a statistical treatment of nuclear-resonance fluorescence spectra~\cite{Rusev} are frequently used.     

For energies below $\sim$3 MeV, the $\gamma$SF of a nucleus is expected to correspond to the exponentially decreasing tail of the giant electric dipole resonance (GEDR). It therefore came as a surprise when a sizable low-energy enhancement in the $\gamma$SF, hereafter referred to as the {\em upbend}, was discovered below 3 MeV for $^{56,57}$Fe~\cite{VoinovLetter}. The $\gamma$SF measurement was performed at the Oslo Cyclotron Laboratory (OCL), using charged particle reactions (and confirmed using the two-step cascade method). In the following years this phenomenon was observed in a wide range of nuclei using the Oslo method~\cite{ACL_Sc, AB_Sc, ACL_Ti, NS_Ti, MG_Ti, ACL_V, MG_Mo}. Recently, the upbend was also reported with a different experimental technique in $^{95}$Mo~\cite{WiedekingLetter}. 

The physical mechanisms behind the upbend have been a puzzle for many years, but intense experimental and theoretical endeavors have recently led to results. Through angular-distribution measurements, it was demonstrated that the upbend is dominantly of dipole nature~\cite{ACL_dipolLetter}. Furthermore, the authors of Ref.~\cite{Litvinova} suggested that the upbend is caused by thermal excitations in the continuum leading to
enhanced low-energy $E$1 transitions. Shell model calculations performed in Refs.~\cite{RSchwengner, ABrown}, on the other hand, show very strong $M$1 transitions at low $\gamma$-ray energies. Moreover, it has been shown~\cite{AC&Goriely} that the presence of the upbend may enhance the r-process (n, $\gamma$) reaction rates by a factor of 10-100. 

Unfortunately, the various experimental techniques based on ($\gamma,\gamma^{\prime}$), ($d,p$), ($^3$He,$^3$He$^{\prime}\gamma$) reactions often give rise to large deviations in the $\gamma$SFs below $S_n$. Therefore, an international collaboration has been formed in order to investigate one specific nucleus as a test case. Germanium-74 was chosen, and four different experiments were performed: ($^3$He,$^3$He$^{\prime}$$\gamma$), ($\alpha$,$\alpha^{\prime}$$\gamma$), (p,p$^{\prime}$$\gamma$) and ($\gamma$,$\gamma^{\prime}$). In this work, we will present results from $^3$He induced reactions on $^{74}$Ge performed at the OCL, and data from a photo-neutron experiment on $^{74}$Ge performed at NewSUBARU in Japan.
In Sec.~\ref{sec:exp}, the experimental details and the data analysis of the two experiments are discussed. The normalization procedure of the OCL data is presented  and a discussion of the resulting $\gamma$SFs is made in Sec.~\ref{sec:normalization}. Shell-model calculations on the $M$1 strength in $^{73,74}$Ge are presented in Sec.~\ref{sec:shell}. Neutron capture cross sections and reaction rates are shown in Sec.~\ref{sec:cross}. Finally, a summary and outlook can be found in Sec.~\ref{sec:conc}. 


\section{Experimental results}
\label{sec:exp}
\subsection{The charged-particle experiment}
\label{sec:oslo_exp}

\begin{figure*}[tb]
\begin{center}
\includegraphics[clip,width=2.\columnwidth]{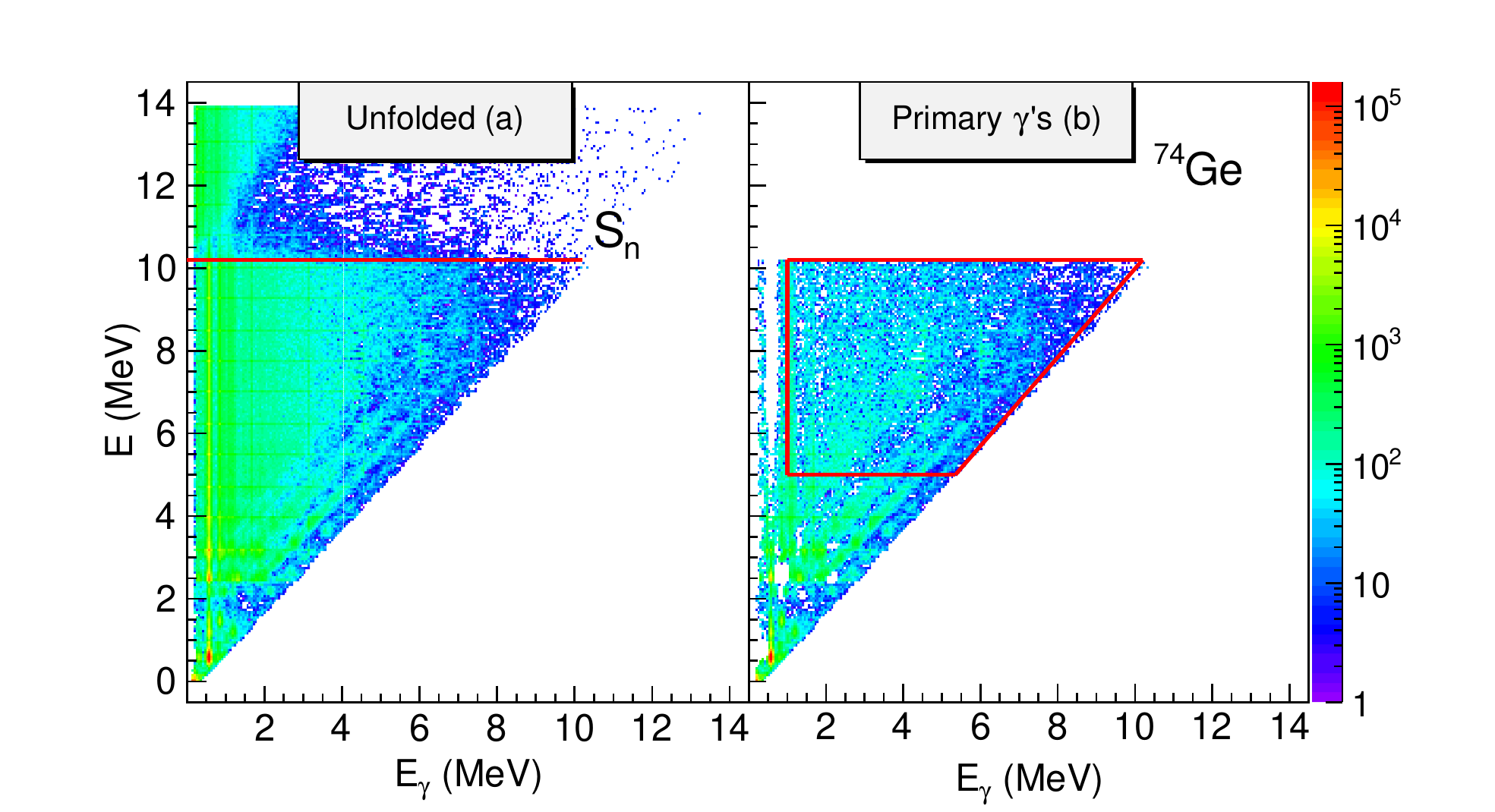}
\caption{(Color online) (a) Excitation energy-$\gamma$ matrix from the $^{74}$Ge($^3$He, $^3$He$\gamma$)$^{74}$Ge reaction. The NaI spectra are unfolded with the NaI response functions. (b) The first generation matrix from the same reaction.} 
\label{fig:Unf&Fg}
\end{center}
\end{figure*}
The charged-particle experiment was performed at the OCL, where a beam of 38-MeV $^3$He particles with a current of $\approx$ 0.5 enA impinged on a self-supporting 0.5 mg/cm$^{2}$ thick $^{74}$Ge target. The target was continuously irradiated for seven days. About 5$\times$10$^6$ and 2$\times$10$^6$ particle-$\gamma$ coincidences were recorded in each of the two reaction channels of interest: $^{74}$Ge($^{3}$He, $^{3}$He$^{\prime}\gamma$) and  $^{74}$Ge($^{3}$He, $\alpha \gamma$). 

The charged outgoing particles were identified and their energies measured with the SiRi system~\cite{SiRi}, consisting of 64 $\Delta$E-E silicon telescopes, with thicknesses of 130-$\mu$m and 1550-$\mu$m, respectively. SiRi was placed in forward direction, covering angles from $\theta$= 40-54$^{\circ}$ and a solid angle coverage of $\approx$6$\%$ of 4$\pi$. The $\gamma$ rays were measured by the CACTUS array~\cite{CACTUS} consisting of 28 collimated 5$^{\prime \prime}$$\times$ 5$^{\prime \prime}$ NaI(Tl) detectors placed on a spherical frame surrounding the target and the particle detectors. The total efficiency of CACTUS is 15.2(1)$\%$ at $E_{\gamma}=1332.5$ keV. 
Using reaction kinematics, the initial excitation energy of the residual nucleus can be deduced from the energy of the outgoing particles detected in SiRi. The particle-$\gamma$ coincidence technique is used to assign each $\gamma$ ray to a cascade depopulating a certain initial excitation energy in the residual nucleus.   
 
Figure~\ref{fig:Unf&Fg}(a) shows the excitation energy-$\gamma$ matrix $(E_\gamma, E)$ of the $^{74}$Ge($^{3}$He, $^{3}$He$^{\prime}$$\gamma$) reaction, where the $\gamma$ spectra have been unfolded~\cite{unfold} with the response functions of CACTUS. The neutron separation energy of $^{74}$Ge is reflected clearly in a drop in $\gamma$ intensity at $E\approx S_n = 10.196$ MeV. A relatively weak diagonal at $E = E_\gamma$, reveals that the direct feeding to the ground state of spin/parity 0$^+$ is not particularly favored in this reaction. A second and third more pronounced diagonal represent direct decay to the 2$^+$ states of 596 keV and 1204 keV, respectively. These $\gamma$ rays stem from primary transitions in the $\gamma$-cascades. 

We would like to study the energy distribution of all primary $\gamma$-rays originating from various excitation energies, and extract level density and $\gamma$SF simultaneously from this information. 
Using the unfolded $(E_\gamma, E)$ matrix, a primary $\gamma$ matrix $P( E_\gamma, E)$, as shown in Fig.~\ref{fig:Unf&Fg}(b), is constructed using the subtraction method of Ref.~\cite{forstegen}. The basic assumption behind this method is that the $\gamma$-ray decay pattern from any excitation bin is independent of whether this bin was populated directly via the ($^{3}$He, $\alpha \gamma$) or ($^{3}$He, $^{3}$He$^{\prime} \gamma$) reactions or indirectly via $\gamma$ decay from higher excitation levels following the initial nuclear reaction. This assumption is fulfilled when states have the same relative probability to be populated by the two processes, since $\gamma$ branching ratios are properties of the levels themselves. 

Fermi's golden rule predicts that the decay probability may be factorized into the transition matrix element between the initial $\ket{i}$ and final states $\bra{f}$, and the density of the final states $\rho _f$~\cite{fermi}:
\begin{equation}
\lambda_{i\rightarrow f} = \frac{2\pi}{\hbar}|\langle f |H^{\prime}| i \rangle|^2\rho_f.
\label{eq:fermi}
\end{equation}
Turning to our first generation $\gamma$-ray spectra $P(E_\gamma, E)$ we realize that they are proportional to the decay probability from $E$ to $E_f$ and we may write the equivalent expression of Eq.~(\ref{eq:fermi}) as:
\begin{equation}
P(E_{\gamma}, E) \propto  {\mathcal{T}}_{{\mathrm{i}}\rightarrow {\mathrm {f}}}\rho,
\label{eq:Oslofermi}
\end{equation}
where ${\mathcal{T}}_{{\mathrm{i}}\rightarrow {\mathrm {f}}}$ is the $\gamma$-ray transmission coefficient, and $\rho = \rho(E-E_{\gamma})$ is the level density at the excitation energy $E_{\mathrm{f}}$ after the first $\gamma$-ray emission. 

We notice that this expression does not allow us to simultaneously extract ${\mathcal{T}}_{{\mathrm{i}}\rightarrow {\mathrm {f}}}$  and $\rho$. To do so, either one of the factorial functions must be known, or some restrictions have to be introduced. Our restriction comes in the form of the Brink-Axel hypothesis~\cite{brink, axel}. The original hypothesis states that the GEDR can be built on any excited state, and that the properties of the GEDR do not depend on the temperature of the nuclear state on which it is built. This hypothesis can be generalized to include not only the GEDR, but any kind of collective nuclear excitation and results in the assumption that primary $\gamma$ spectra originating from the excitation energy $E$ can be factorized into a $\gamma$-ray transmission coefficient ${\mathcal{T}(E_\gamma)}$, which depends only on the $\gamma$-transition energy $E_{\gamma}$, and into the level density $\rho(E-E_{\gamma})$ at the final level. We have now the simple relation:
\begin{equation}
P(E_{\gamma},E) \propto  {\mathcal{T}}(E_{\gamma}) \rho(E-E_{\gamma}),
\label{eq:Oslofermi}
\end{equation}
which permits a simultaneous extraction of the two functions from the first-generation matrix. At low excitation energies, the $\gamma$ decay is, naturally, highly dependent on the individual initial and final state. This has been taken into consideration in our analysis, and we have excluded the $\gamma$-ray spectra originating from excitation energy bins below 3 MeV for $^{73}$Ge and 5 MeV for $^{74}$Ge. Also, a lower limit is set on the $\gamma$ rays, where $E_{\gamma}^{min}$ is 1 MeV and 1.5 MeV for $^{73}$Ge and $^{74}$Ge, respectively. In the range of  $E_{\gamma} < E_{\gamma}^{min}$, strong, discrete transitions are too heavily or to modestly subtracted in the first generation method, and are thus excluded from further analysis. 

At this point in the analysis, we have established the functional form of the level density and transmission coefficient. As demonstrated in Ref.~\cite{schiller_geni}, there exists an infinite set of solutions to Eq.~(\ref{eq:Oslofermi}), using transformations. The last stages of the analysis of the OCL data and the normalization procedure will be described in Sec.~\ref{sec:normalization}.      


\subsection{The photo-neutron experiment}
\label{sec:japan_exp}

\begin{figure}[b]
 \begin{center}
\includegraphics[clip,width=8 cm]{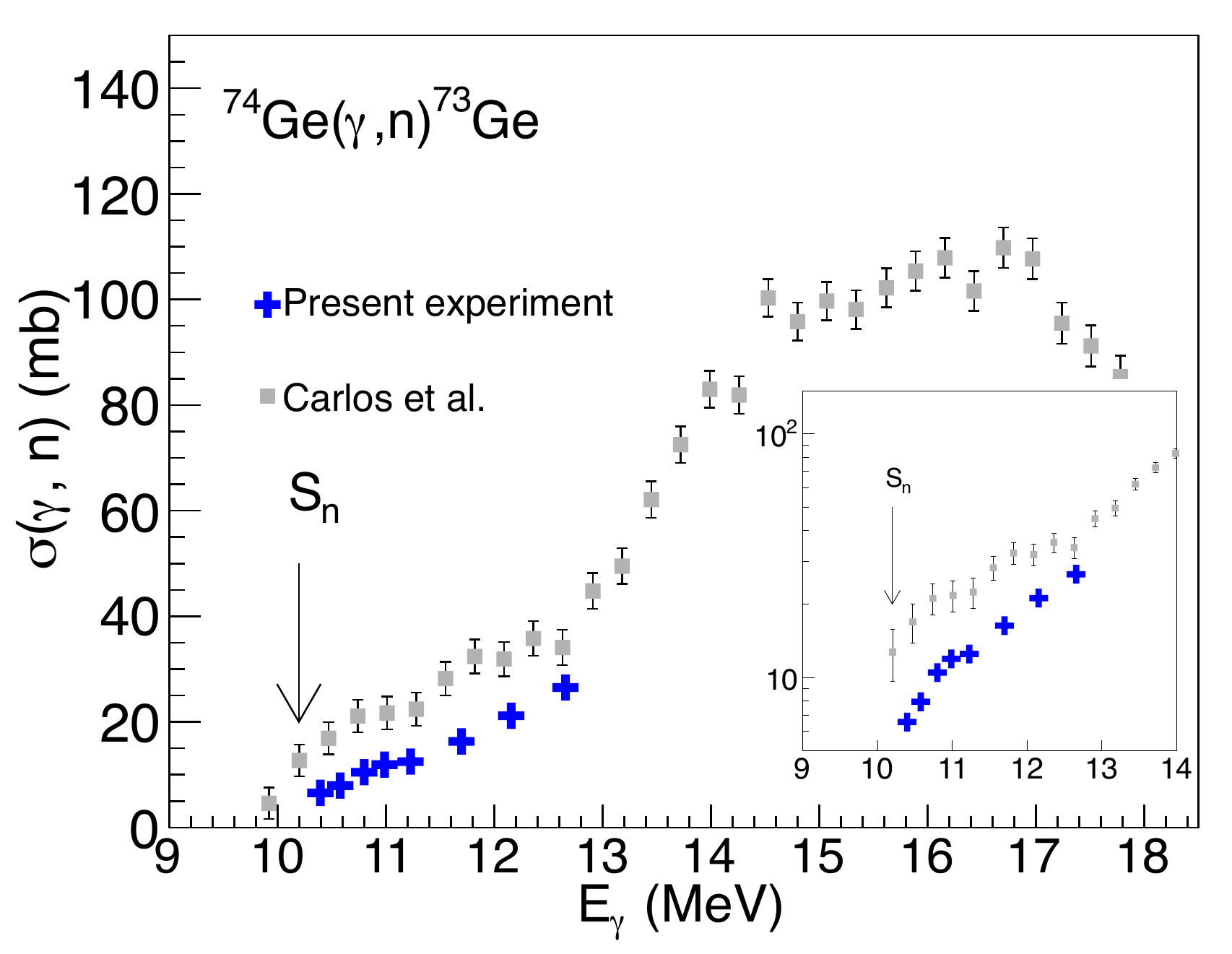}
 \caption {(Color online) Cross sections for the $^{74}$Ge($\gamma$, n) reaction from the current experiment (blue crosses) together with existing photo-neutron data ~\cite{carlos}.}
 \label{fig:cross74Ge}
 \end{center}
 \end{figure}

The photo-neutron cross section measurement was performed at the synchrotron radiation facility NewSUBARU in the Hy$\bar{\mathrm{o}}$go Prefecture~\cite{newsubaru}. Here a wide range of quasi monochromatic $\gamma$ beams~\cite{newsubaru2} are produced in head-on collisions between laser photons and relativistic electrons, so-called laser Compton scattering (LCS). The energy of the laser photons increases from a few eV to several MeV in the collision. In this experiment a 1.99 g/cm$^2$ thick sample of $^{74}$Ge, enriched to 97.53$\%$, was placed inside an aluminum container and irradiated with eight different $\gamma$ beams with energies ranging from 10.4 to 12.7 MeV. 

The $^{74}$Ge sample was mounted in the center of a 4$\pi$ neutron detection array comprised of 20 $^3$He proportional counters embedded in a 36$\times$36$\times$50~cm$^3$ polyethylene moderator. The ring ratio technique~\cite{originalRR} was used to measure the average energies of the detected neutrons, and from this we establish the efficiency of the neutron detector as a function of neutron energy. A 6$^{\prime\prime}\times$5$^{\prime\prime}$ NaI(Tl) detector was used to measure the flux of the LCS beam. The detector was placed at the end of the $\gamma$-ray beam line. The intensity of $\gamma$ rays hitting the $^{74}$Ge target was $\approx 10^5$ s$^{-1}$. The total number of $\gamma$ rays on target for a certain beam energy was found using the pile-up method described in Ref.~\cite{kondo}. The almost monochromatic $\gamma$ beams were monitored by a 3.5$^{\prime\prime}\times$4.0$^{\prime\prime}$ LaBr$_{3}$(Ce) detector between every neutron measurement run. These spectra are reproduced using GEANT4~\cite{GEANT4} simulations, and unfolded to extract the real energy profile of the incoming beam. 

The ($\gamma$, n) cross section is given by
\begin{equation}
\int_{S_n}^{E_{\rm{Max}}}n_{\gamma}(E_{\gamma})\sigma(E_{\gamma})dE_{\gamma}=\frac{N_n}{N_tN_{\gamma}\xi\epsilon_n g},
\label{eq:mono1}
\end{equation}
where $n_{\gamma}(E_{\gamma})$ denotes the energy distribution of the $\gamma$-ray beam normalized to unity, and $\sigma(E_{\gamma})$ is the photo-neutron cross section to be determined. Furthermore, $N_n$ represents the number of neutrons detected, $N_t$ gives the number of target nuclei per unit area, $N_{\gamma}$ is the number of $\gamma$ rays incident on target, $\epsilon_n$ represents the neutron detection efficiency, and $\xi=(1-e^{\mu t})/(\mu t)$ gives a correction factor for a thick target measurement, where $t$ is the thickness of the target and $\mu$ is the attenuation coefficient of the target. The factor $g$ represents the fraction of the $\gamma$ flux above $S_n$.
Equation (\ref{eq:mono1}) is solved for the cross section using a Taylor expansion method described in Ref.~\cite{DanIoana_Sm}. In this way, we find cross sections for eight different energies, starting from 200 keV above S$_n$ of $^{74}$Ge. The total uncertainties in the measurements are $\approx$4.4$ \%$~\cite{HildeTherese_Nd}. The resulting $^{74}$Ge($\gamma$,n) cross sections are shown in Fig.~\ref{fig:cross74Ge}. We note that the newly measured data are lower than the data retrieved from a positron annihilation in-flight experiment by Carlos {\it et al.}~\cite{carlos} by $\approx$ 30 $\%$. The same trend has been reported by Berman {\it et al.}~\cite{Berman_lav}. In the insert in Fig.~\ref{fig:cross74Ge}, the difference in shape between the two datasets becomes more apparent; our newly measured cross sections vanish at $\approx S_n$ as expected, whereas the Carlos data exhibit a non-zero value in this range.   

\section{Normalization of the OCL data}
\label{sec:normalization}
Once we have extracted the two vectors $\rho(E-E_{\gamma})$ and $\mathcal T(E_{\gamma})$ from the first generation matrix, we can construct infinitely many solutions~\cite{schiller_geni} that give identical fits to the experimental data. The set of solutions are of the form
\begin{align}
\tilde{\rho}(E-E_\gamma)&={\rm{A}}\exp[\alpha(E-E_\gamma)]\,\rho(E-E_\gamma),\label{eq:array0}\\
\tilde{{\mathcal{T}}}(E_\gamma)&={\rm{B}}\exp(\alpha E_\gamma){\mathcal{T}} (E_\gamma),\label{eq:array1}
\end{align}
and it is necessary to determine the transformation coefficients A, $\alpha$ and B that gives solutions corresponding to the actual level densities and $\gamma$-transmission coefficients of $^{73,74}$Ge. To be able to do this, we take advantage of auxiliary data, mainly stemming from neutron resonance experiments. This process of determining the coefficients and thus the physical solutions is what we refer to as \emph{normalization} of our experimental data.


\subsection{Level density}
\begin{figure}[b]
\begin{center}
\includegraphics[clip,width=8cm]{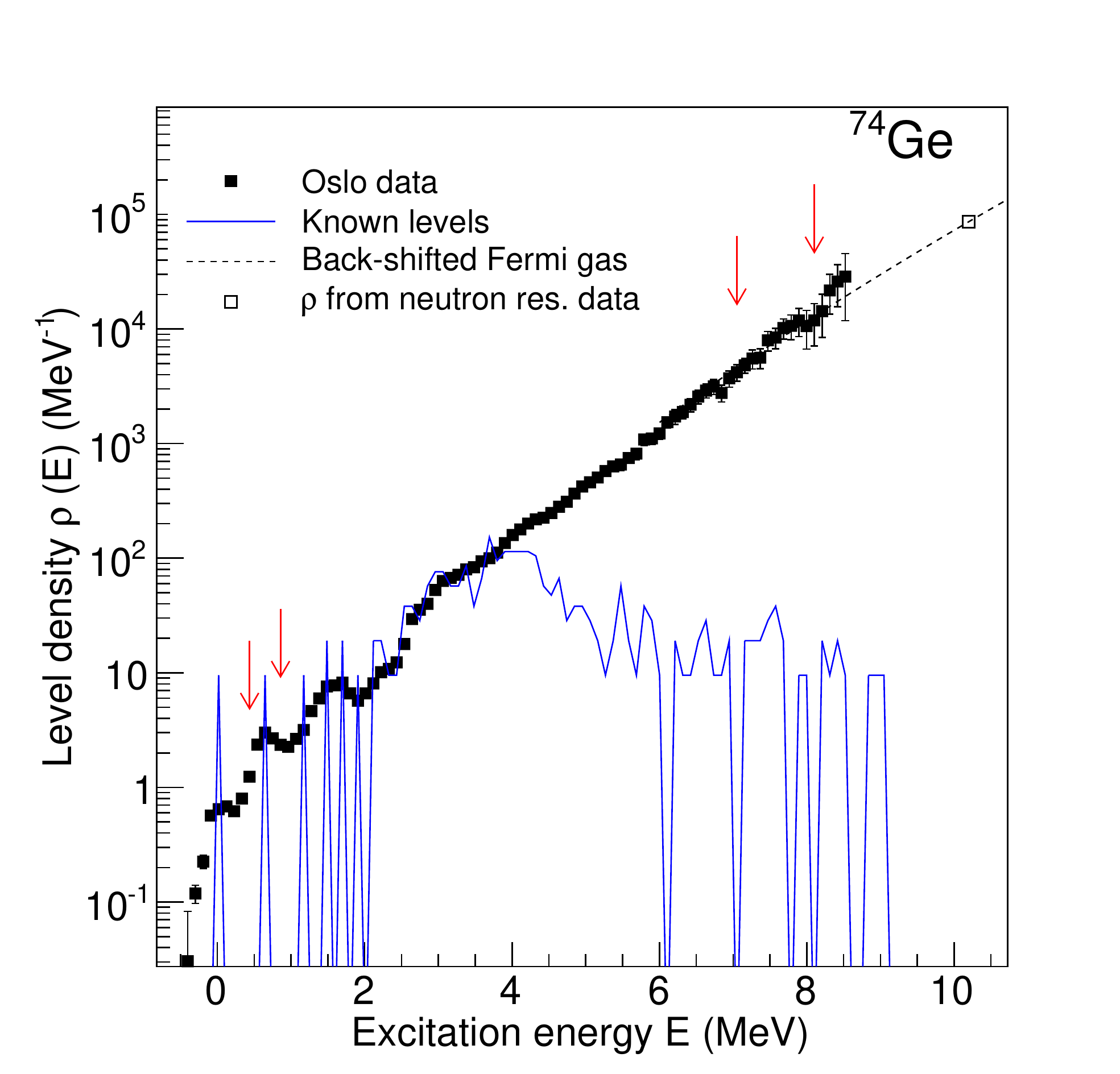}
\caption{(Color online) Experimental level densitiy of $^{74}$Ge. The data are normalized to known discrete levels at low excitation energy and to the level density extracted at $S_n$ from neutron capture resonance spacings $D_0$. The two set of arrows indicate where the data are normalized.}
\label{fig:norm_levels}
\end{center}
\end{figure}

We start by establishing the normalized nuclear level densities (NLD) of $^{73, 74}$Ge. This entails determining the two coefficients A and $\alpha$ of Eq.~(\ref{eq:array0}). For this purpose we need two anchor points, i.e. two regions of excitation energy where there exist information on the NLD, either from experimental data or from theoretical calculations. A proper spacing between the anchor points is essential to ensure a reliable normalization. At low excitation energies, known, discrete levels can be used. The anchor points at low excitation energies of the two Ge isotopes are found simply by using the definition of NLD, $\rho = \frac{\Delta N}{\Delta E}$, where $\Delta N$ is the number of levels in the $\Delta E$ energy bin, using the same bin size as our experimental one, where $\Delta E$ = 105 keV. The level schemes of $^{73,74}$Ge are assumed to be close to complete up to excitation energies of 1.38 MeV and 3.4 MeV, respectively~\cite{NNDC}, and we choose an area below these energies for normalization, see Fig.~\ref{fig:norm_levels}, where the arrows to the left show the area used in the case of $^{74}$Ge. 

The second anchor point is at higher energies, where the most reliable experimental data on the NLD comes from neutron resonance experiments that provide average neutron spacings $D$ in the area of the neutron separation energy. In the case of $^{73,74}$Ge, $s$-wave spacings, $D_0$, are given in both RIPL-3~\cite{RIPL3} and the {\it Atlas~of~Neutron~Resonances} of S.~F.~Mughabghab~\cite{Mug}, from neutron capture on $^{72}$Ge and $^{73}$Ge. After careful consideration, we have chosen to use an average value of the two proposed sets of $D_0$ values and uncertainties. The two main reasons for this choice are: 
\begin{enumerate}
\item For $^{73,74}$Ge the $D_0$ values from Ref.~\cite{Mug} are larger by 38$\%$ and 60$\%$, respectively, than the values given in Ref.~\cite{RIPL3}. 

\item Ref.~\cite{Mug} presents a table of measured resonances. The experimental results that give the values listed in Ref.~\cite{RIPL3} are, to our knowledge, not presented in any peer-reviewed publication.
\end{enumerate}

The s-wave spacings can be expressed in terms of partial level densities:
\begin{align}
D_0 &= \frac{1}{\rho(S_n, I_t+1/2, \pi_t) + \rho(S_n, I_t-1/2, \pi_t) }  \\ 
         &  \hspace{3.0cm }\text{for}\ I_t >0,  \\                                                               
         &= \frac{1}{\rho(S_n, 1/2, \pi_t)} \hspace{0.55cm } \text{for} \ I_t = 0 ,
\label{eq:D0}
\end{align}
where $I_t$ and $\pi_t$ are the spin and parity, respectively, of the target nucleus. From Eq.~(\ref{eq:D0}), we find that the measured level spacing $D_0$ in the case of the $^{72}$Ge(n,$\gamma$) reaction corresponds to the density of $\frac{1}{2}+$ states in $^{73}$Ge at $S_n$ = 6.783 MeV, $\rho_{\frac{1}{2}^{+}}(S_n)$. From $^{73}$Ge(n, $\gamma$), the density of $4^{+}$ and $5^{+}$ states of $^{74}$Ge at $S_n$ = 10.196 MeV, $\rho_{4^+,5^+}(S_n)$, can be estimated. Our experimental NLD represents the density of almost all accessible spins at $S_n$. From semi-classical calculations, we get $I_{max} \simeq$ 10 $\hbar$ for a $^3$He beam at 38 MeV.
Due to the lower limits applied on E$_\gamma$ (see Sec.~\ref{sec:exp}) in the extraction of $\rho$ and $\mathcal T$, our NLDs reach only up to E $\approx$ $S_n$ - 1.5 MeV. We need to make an interpolation between our data points and the NLD at S$_n$. The back-shifted Fermi gas model with the parameterization of Egidy and Bucurescu~\cite{E&B2009} has been chosen for this purpose (see Tab.~\ref{table:level_parameters}). Another option had been to use a contant temperature model for the interpolation as recommended in Ref.~\cite{Guttormsen_level}, but in this case where the gap between the last data point and $S_n$ is so small the two types of interpolations gives very similar results (see Ref.~\cite{Toft_Sn}).
From this point in the analysis we will carry out the normalization according to two different normalization schemes.

\subsubsection{norm-1}
The main idea of this approach is to go from the spin- and parity-dependent NLD to the total NLD at $S_n$:
\begin{equation}
\rho_{tot}(S_n) = \sum_I \sum_{\pi} \rho(S_n, I, \pi).
\label{eq:EJpi_rho}
\end{equation}
This equation shows that we need information about the spin and parity distribution around the neutron separation energy. These quantities are both notoriously difficult to measure experimentally for all spins and both parities at such high excitation energies. 
At this point two assumptions are made:
\begin{enumerate}
\item The spin dependence of the level density is given by the statistical approximation~\cite{Ericson1, Ericson2}:
\begin{equation}
 g(E, I) \simeq \frac{2I + 1}{2\sigma^2}\rm{exp} [-(I+1/2)^2/2\sigma^2 ], 
 \label{eq:G&C_spin}
 \end{equation}
where $I$ is the spin. The spin cut off parameter, $\sigma$, is parameterized as recommended in Ref.~\cite{E&B2009}. The fact that in the case of $^{74}$Ge we know the density of 4$^+$ and 5$^+$ states that lie close to the center of the assumed spin distribution, will most probably lead to a good estimation of the full spin distribution. 
\item There is an equipartition of parities at the neutron separation energy for the two Ge isotopes. The assumption of parity symmetry at these high excitation energies for nuclei in this mass region is supported by~\cite{Kalmykov_parity}. 
\end{enumerate}
We can now express the level density at $S_n$ by~\cite{schiller_geni}:
\begin{gather}
\rho(S_n) = \frac{2\sigma^2}{D_0} \frac{1}{(I_{t} + 1) \rm{exp}[-(I_{t}+ 1)^2/2\sigma^2 ] + I_{t} \rm{exp}[-I_{t}^2/2\sigma^2]},
\raisetag{-0.5em}
\label{eq:level_Sn}
\end{gather}
and we have found our second anchor point.

\subsubsection{norm-2}
Recent microscopic calculations~\cite{Goriely_level2008, Hilarie_level2006, Hilarie_level2001} based on the Hartree-Fock-Bogolyuobov (HFB) plus combinatorial (HFB + Comb.) approach have been successful in calculating NLDs of a wide range of nuclei.  
Such an approach provides the energy, spin and parity dependence of the NLD. For flexibility, the calculated NLD can be normalized, if need be, to reproduce experimental $s$-wave spacings and the density of discrete levels at low excitation energies. These calculations have no a priori assumptions on the spin or parity distribution. The microscopic calculations generally give a broad spin distribution with a center of gravity at quite high spins, and provide a higher NLD at $S_n$ (see Tab.~\ref{table:level_parameters}).
   
We keep in mind that the different normalizations, {\it norm-1} and {\it norm-2}, will lead to different slopes, $\alpha$, of the normalized level density, and because of their interconnection also determine the slope for the $\gamma$-transmission coefficient, see Eqs.~(\ref{eq:array0}) and ~(\ref{eq:array1}).

\begin{table*}
\caption{Parameters used in the normalization of NLD and $\gamma$SF}
\centering
\begin{tabular}{c c c c c c c c c c}
\hline\hline
Nucleus & $I_t^{\pi}$        & $D_0$            &   $S_n$     & $\sigma(S_n)$&  a                 & $E_1$       & $\rho(S_n)_{\mathrm{norm-1}}$      &    $\rho(S_n)_{\mathrm{norm-2}}$      &   $\left<\Gamma_{\gamma}\right>$              \\ 
                &                    & (eV)            & (MeV)          &                            & (MeV$^{-1}$) & (MeV)    &      (MeV$^{-1}$)                     &     (MeV$^{-1}$)               &              (meV)                                     \\ 
\hline
$^{73}$Ge &     0$^+$        &1785(209)  &  6.783          &  3.66                  &        9.00      &  -1.32    &      156(35) $\cdot$ $10^2$                 &       23521                 &              195(50)                                 \\
$^{74}$Ge &    9/2$^+$      &80.5(9)        & 10.196         &  3.77                  &        9.70      &    0.71   &       860(98) $\cdot$ $10^2$                &       98083                 &              196(23)                                  \\
\hline
\end{tabular}
\label{table:level_parameters}
\end{table*}

\begin{figure}
\begin{center}
\includegraphics[clip,width=\columnwidth]{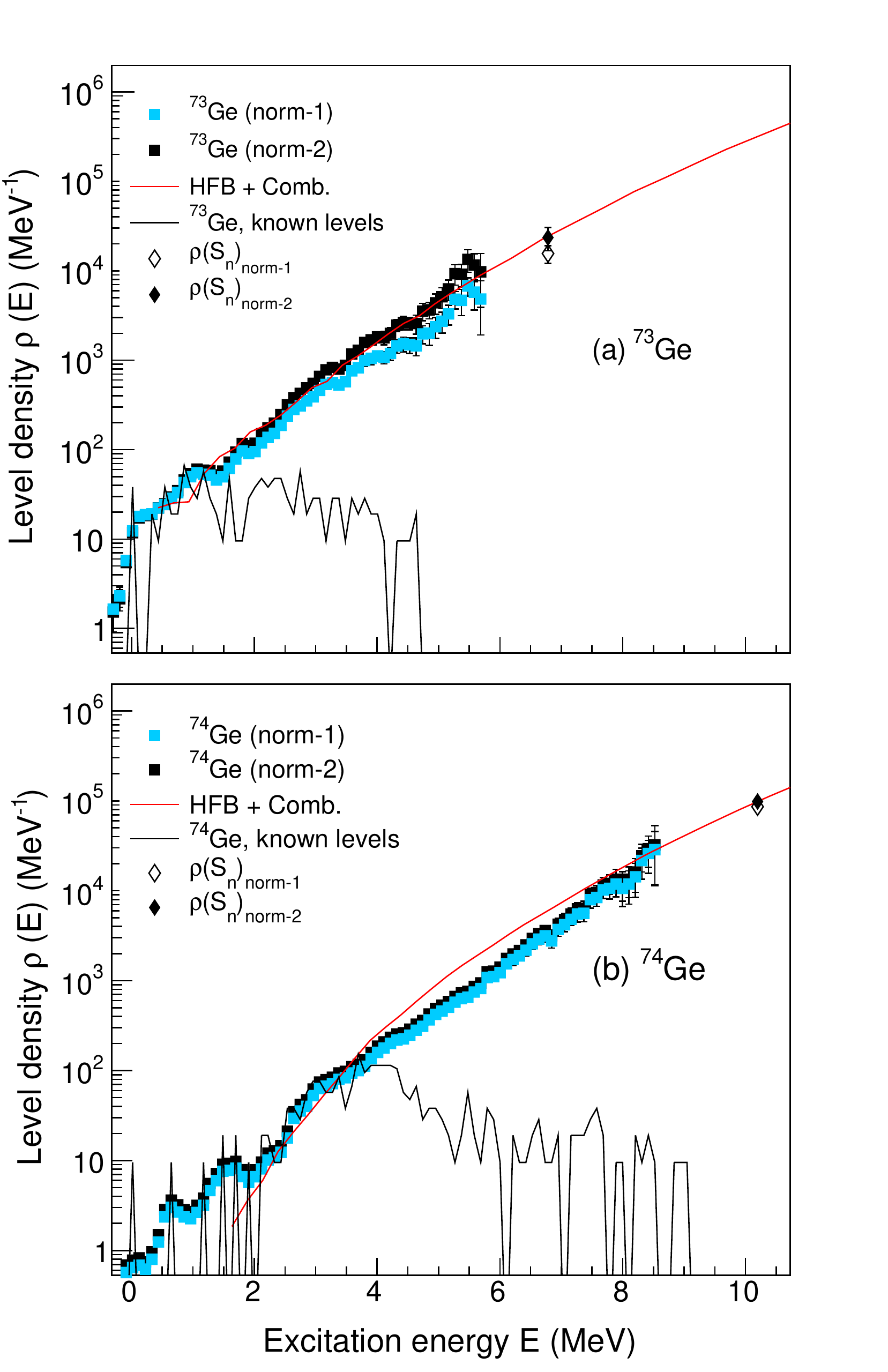}
\caption{(Color online) Level density of (a) $^{73}$Ge and (b) $^{74}$Ge normalized according to {\it norm-1} and {\it norm-2}.}
\label{fig:microfeno}
\end{center}
\end{figure}

In Fig.~\ref{fig:microfeno} the results of the two normalizations of the NLDs are presented. \emph{Norm-1} and \emph{norm-2} give a lower and upper limit of the normalization of the two NLDs. We see that the discrete levels at low excitation energy are well reproduced, and that the good statistics of the experiment yields small statistical errors. 

It is seen that the unpaired neutron in $^{73}$Ge implies a higher NLD than for $^{74}$Ge. Another striking feature is the linearity of the NLD in log scale. This means that the NLD can be well described by the constant temperature expression: 
\begin{equation}
\rho_{\mathrm{CT}}(E) = \frac{1}{T_{\mathrm{CT}}} \mathrm{exp} \frac{(E-E_0)}{T_{\mathrm{CT}}},
\end{equation}
where $T_{CT}$ is determined by the slope of ln$\rho$(E). This linearity has previously been observed for many nuclei, and is described in detail in~\cite{Guttormsen_level, Moretto}. 



\subsection{$\gamma$-transmission coefficient and $\gamma$SF}
The normalization of the $\gamma$-transmisson coefficient, $\mathcal T(E_{\gamma})$, consists of determining the scaling factor B in Eq.~(\ref{eq:array1}) as $\alpha$ is already determined. Average radiative widths of neutron resonances $\left< \Gamma_{\gamma}\right>$ are very important properties of $\gamma$-decay from nuclear states at high excitation energy, and can be used to normalize $\mathcal T(E_{\gamma})$. We normalize according to~\cite{AVoinovGamma}:
\begin{align}
\langle \Gamma_{\gamma}(S_n,I_{t}\pm 1/2,&\pi_t)\rangle =
 \frac{D_0}{4\pi}\int_{E_{\gamma}=0}^{S_n}\mathrm{d}E_{\gamma} B \mathcal{T}(E_{\gamma}) \nonumber \\ 
 &\times \rho(S_n-E_{\gamma}) \sum_{I= -1}^{1} g(S_{n}-E_{\gamma},I_{t}\pm 1/2+I),
\label{eq:width}
\end{align}
where $I_t$ and $\pi_t$ are the spin and parity of the target nucleus in the $(n,\gamma)$ reaction, and $\rho(S_n-E_{\gamma})$ is the experimental level density.

The total average radiative widths are rather complex, depending on the $\gamma$-transmission coefficient, the NLD and the spin distribution. An average of the listed experimental values of $\left <\Gamma_{\gamma}\right >$ from Refs.~\cite{Mug, RIPL3} is taken for $^{74}$Ge, giving $\left<\Gamma_{\gamma}\right>$ = 196(9) meV. The large number of resonances listed in~\cite{Mug} and gives us confidence in the quite low uncertainty in this quantity. We also note that this value of  $\left <\Gamma_{\gamma}\right>$ gives a good agreement with the newly measured ($\gamma,n$) data. Concerning the listed value of $\left<\Gamma_{\gamma}\right>$ for $^{73}$Ge in~\cite{Mug}, we notice that this average value is only based on four experimental values, ranging from between 130 and 200 meV, giving an average value of 150(35) meV. RIPL3\cite{RIPL3} lists a value of 162(50) meV. These values are 15-30 $\%$ lower than the average value for $^{74}$Ge. Considering the poor statistics of $^{73}$Ge compared to the case of $^{74}$Ge, we have chosen to set the $\left <\Gamma_{\gamma}\right>$ value of $^{73}$Ge to 195(50) meV, also in order to be consistent with the ($\gamma$,n) data for $^{74}$Ge. As for the NLDs, we follow here two parallel normalization schemes.
Lastly, taking into account that the transitions between states in the quasi continuum are dominantly of dipole type (e.g~\cite{JKopecky, ACL_dipolLetter} ) the $\gamma$-transmission coefficient, $\mathcal T (E_{\gamma})$, relates to the $\gamma$SF, $f(E_\gamma)$, in the following way~\cite{RIPL3}:
\begin{equation}
f(E_{\gamma}) = \frac{\mathcal T(E_{\gamma})}{2\pi E_{\gamma}^3}.
\end{equation}
We thus deduce the dipole strength from the normalized $\gamma$-transmission coefficient.

Coming back to the photo-neutron cross sections, they are related to the $\gamma$SF, $f(E_{\gamma})$, by
\begin{equation}
f(E_{\gamma})= \frac{1}{3\pi^2\hbar^2c^2} \frac{\sigma(E_{\gamma})}{E_{\gamma}},
\label{eq:cross_toRSF}
\end{equation}
which can be directly compared with the Oslo data from the principle of detailed balance, giving $f_{\mathrm{up}} \approx f_{\mathrm{down}}$~\cite{RIPL3}.

\begin{figure}
\begin{center}
\includegraphics[clip,width=\columnwidth]{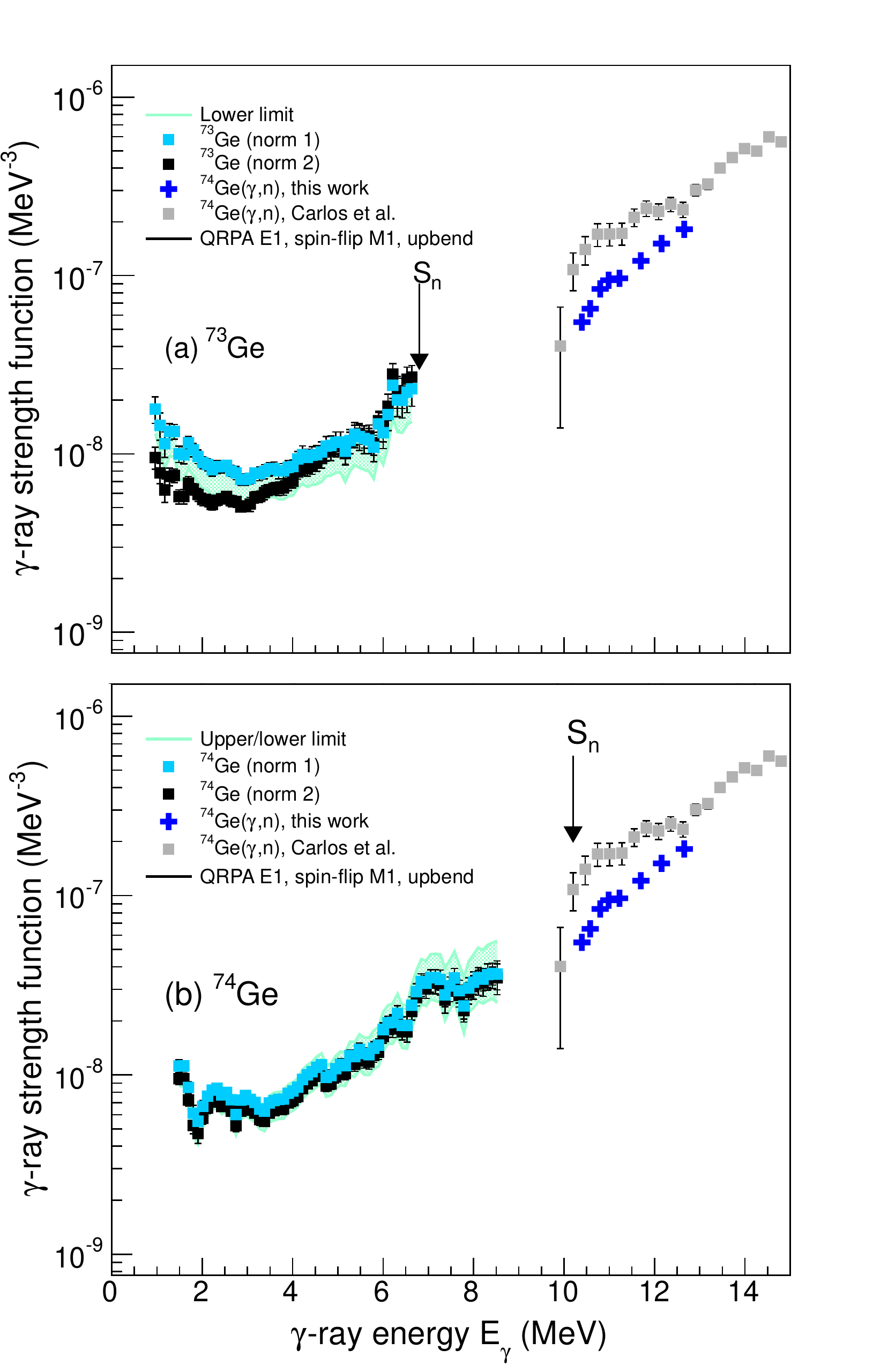}
\caption{(Color online) $\gamma$SF for the different normalization procedures together with photo-neutron data on $^{74}$Ge from the present experiment and from already existing photo-neutron data from Ref.~\cite{carlos} for (a) $^{73}$Ge and (b) $^{74}$Ge. Note that $^{74}$Ge $\gamma$SF above $S_n$ has been compared with both $^{73,74}$Ge data from the current experiment. The green lines represent the upper and lower limits of {\it norm-1}.}
\label{fig:level73}
\end{center}
\end{figure}

Now we are ready to present the $\gamma$SF below $S_n$ together with the data points above $S_n$ from the photo-neutron experiment. The $\gamma$SFs of $^{73,74}$Ge from the two normalization methods, \emph{norm-1} and \emph{norm-2}, are shown in Fig.~\ref{fig:level73}. The error bars of the data points include statistical errors, and propagated systematic errors from the unfolding and the primary $\gamma$-ray extraction. Systematic errors originating from the normalization process are indicated as upper and lower limits. We notice that \emph{norm-2} gives lower and steeper $\gamma$SFs than \emph{norm-1} in the case of $^{73}$Ge, but in the case of $^{74}$Ge the two normalization schemes give very similar results. The $\gamma$SFs below $S_n$ are in both cases in good agreement with the new photo-neutron data on $^{74}$Ge. 

We observe a resonance-like structure centered at $\approx$ 7 MeV. This has also been observed in the $^{74}$Ge($\alpha$,$\alpha^{\prime}\gamma$)$^{74}$Ge reaction~\cite{MWiedekingPrivate}. Strength functions from the $^{74}$Ge($\gamma,gamma^{\prime}$) experiment are in good agreement with the results presented here~\cite{gamma_gamma}. We also see that both the $\gamma$SFs of $^{73,74}$Ge are increasing at decreasing $\gamma$-ray energies below $\sim$ 3 MeV. This finding is expected from the results of Ref.~\cite{AC_76Ge}, where the $\gamma$SF of $^{76}$Ge was reported to show a similar upbend. In the following, we will compare our data with calculations of the $M1$ strength. 

\section{Shell-model calculations of the $M$1 strength}
\label{sec:shell}

We have performed shell-model calculations by means of the code RITSSCHIL~\cite{zwa85} using a model space composed of the
$(0f_{5/2}, 1p_{3/2}, 1p_{1/2}, 0g_{9/2})$ proton orbits and the
$(1p_{1/2}, 0g_{9/2}, 1d_{5/2})$ neutron orbits relative to a $^{66}$Ni
core. This configuration space is analogous to the one applied in an earlier
study of $M1$ strength functions in $^{94,95,96}$Mo and $^{90}$Zr~\cite{RSchwengner}.
In the present calculations for $^{73,74}$Ge, four protons were allowed to be
lifted from the $(fp)$ shell to the $0g_{9/2}$ orbit and two neutrons from the
$1p_{1/2}$ to the $0g_{9/2}$ orbit. This resulted in dimensions up to 11400. For
comparison, $M1$ strength functions were deduced also for the neutron-rich isotope $^{80}$Ge.
In these calculations, one neutron could be excited from the $0g_{9/2}$ to the
$1d_{5/2}$ orbit. We note here that the resctricted model space does not fully reproduce the collectivity in the near-yrast states of $^{74}$Ge. However, the calculations give an approach to the characteristics of $M1$ transitions between excited states above the yrast line~\cite{RSchwengner,sch14E2}.

The calculations included states with spins from $I$ = 0 to 10. For each spin
the lowest 40 states were calculated. Reduced transition probabilities $B(M1)$
were calculated for all transitions from initial to final states with
energies $E_f < E_i$ and spins $I_f = I_i, I_i \pm 1$. For the minimum and
maximum $I_i$, the cases $I_f = I_i - 1$ and $I_f = I_i + 1$, respectively,
were excluded. This resulted in more than 23800 $M1$ transitions for each
parity $\pi = +$ and $\pi = -$, which were sorted into 100 keV bins according
to their transition energy $E_\gamma = E_i - E_f$. The average $B(M1)$ value
for one energy bin was obtained as the sum of all $B(M1)$ values divided by the
number of transitions within this bin.

\begin{figure}
\begin{center}
\includegraphics[clip,height=6.5cm]{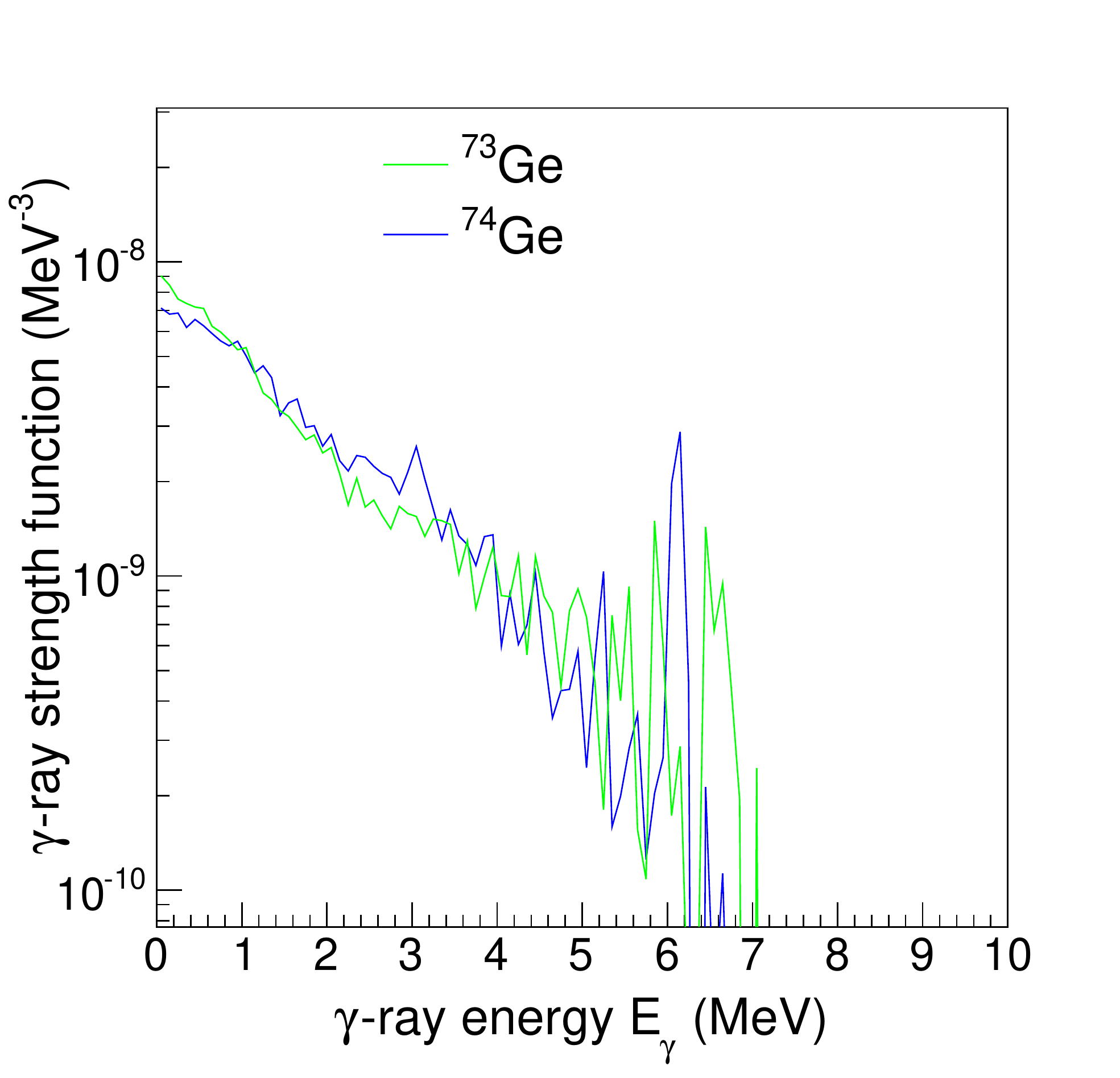}
\caption{(Color online) Shell model calculations of the $M1$ component of the $\gamma$SF of $^{74}$Ge (blue line) and $^{73}$Ge (green line).}
\label{fig:74Ge-sf}
\end{center}
\end{figure}

The $M1$ strength functions were deduced using the relation
\begin{equation}
f_{M1}(E_\gamma)
= 16\pi/9 (\hbar c)^{-3} \overline{B}(M1,E_\gamma) \rho(E_i).
\end{equation}
They were calculated by multiplying the ${B(M1)}$ value in $\mu^2_N$ of each
transition with $11.5473 \times 10^{-9}$ times the level density at the energy
of the initial state $\rho(E_i)$ in MeV$^{-1}$ and deducing averages in energy
bins as done for the $\overline{B}(M1)$ values (see above). The level densities
$\rho(E_i,\pi)$ were determined by counting the calculated levels within energy
intervals of 1 MeV for the two parities separately. The strength functions
obtained for the two parities were subsequently added. Gates were put on the excitation energy E$_{x}$, corresponding to the ones
applied in the analysis of the experimental data (see Sec.~\ref{sec:exp}). 
The resulting $M1$ strength function for $^{74}$Ge is shown in
Fig.~\ref{fig:74Ge-sf}. 

\begin{figure}
\begin{center}
\includegraphics[clip,height=6.5cm]{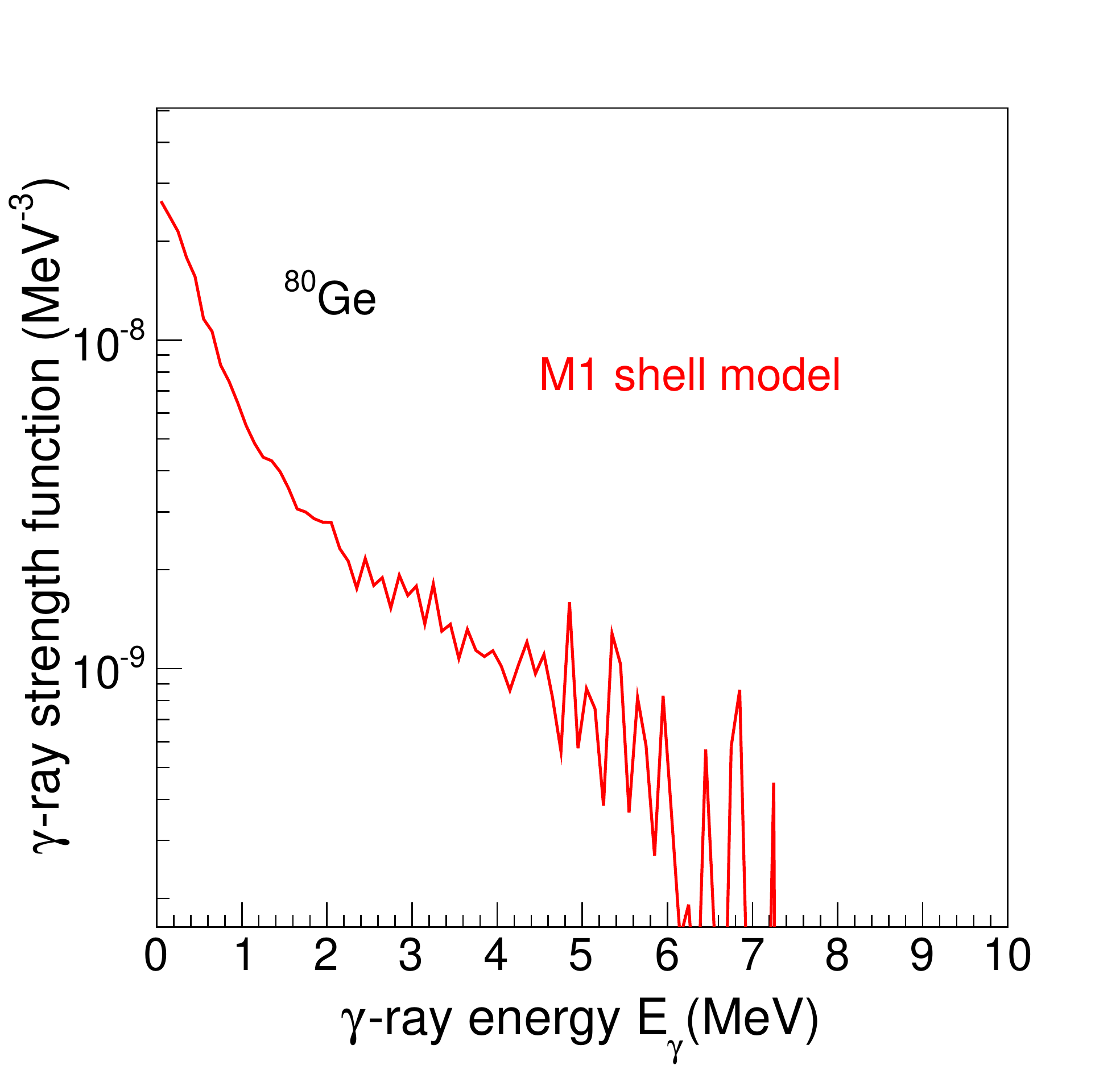}
\caption{(Color online) Shell model calculations of the $M1$ component of the $\gamma$SF of $^{80}$Ge (red line) .}
\label{fig:80Ge-sf}
\end{center}
\end{figure}

The calculated $M1$ strength function shows a low-energy
enhancement similar to that of the $M1$ strength functions calculated for the
neighboring nuclei $^{94,95,96}$Mo, $^{90}$Zr~\cite{RSchwengner} and for
$^{56,57}$Fe~\cite{ABrown}. However, the slope is not as steep as in the nuclides
close to $N$ = 50~\cite{RSchwengner}. The $M1$ strength function calculated for
the $N$ = 48 isotope $^{80}$Ge is shown in Fig.~\ref{fig:80Ge-sf}. One sees that the slope of this
is steeper than that of $^{74}$Ge and reaches larger values toward
$E_\gamma$ = 0. 
The dominating configurations of states in $^{73}$Ge linked by transitions with
large $B(M1)$ values are of the type $\pi[(0f_{5/2} 1p_{3/2})^4]$ $\nu(1p_{1/2}^{-2} 0g_{9/2}^3)$ for positive-parity
states and $\pi[(0f_{5/2} 1p_{3/2})^4]$ $\nu(1p_{1/2}^{-1} 0g_{9/2}^2)$ for negative-parity states. In $^{74}$Ge, the configurations are analogous,
including one $0g_{9/2}$ neutron more. In addition, configurations of the type $\pi[(0f_{5/2} 1p_{3/2})^4]$ $\nu(0g_{9/2}^{2})$ contribute for positive parity.
The corresponding configurations in $^{80}$Ge are $\pi[(0f_{5/2} 1p_{3/2})^4]$ $\nu(0g_{9/2}^8)$ and
$\pi[(0f_{5/2} 1p_{3/2})^4]$ $\nu(0g_{9/2}^7 1d_{5/2}^1)$ for positive-parity states, and $\pi[(0f_{5/2} 1p_{3/2})^3 0g_{9/2}^1]$ $\nu(0g_{9/2}^8)$ and 
$\pi[(0f_{5/2} 1p_{3/2})^3 0g_{9/2}^1]$ $\nu(0g_{9/2}^7 1d_{5/2}^1)$ for negative-parity states.

\section{Calculations of (\lowercase{n},$\gamma$) cross section and reaction rates}
\label{sec:cross}

The measured NLD and the $\gamma$SF together with the nucleon-nucleus optical potential, assuming a compound reaction, can now be applied to calculate the neutron-capture cross section. It has been shown that using the experimental NLD  and $\gamma$SF extracted using the Oslo method as input for (n, $\gamma$) cross section calculations gives a very good agreement with experimental cross section data~\cite{Tamas}. In this work, we focus on the cross sections $^{73}$Ge(n, $\gamma$)$^{74}$Ge and $^{72}$Ge(n, $\gamma$)$^{73}$Ge. 

\begin{figure}
\begin{center}
\includegraphics[clip, height=5.9cm]{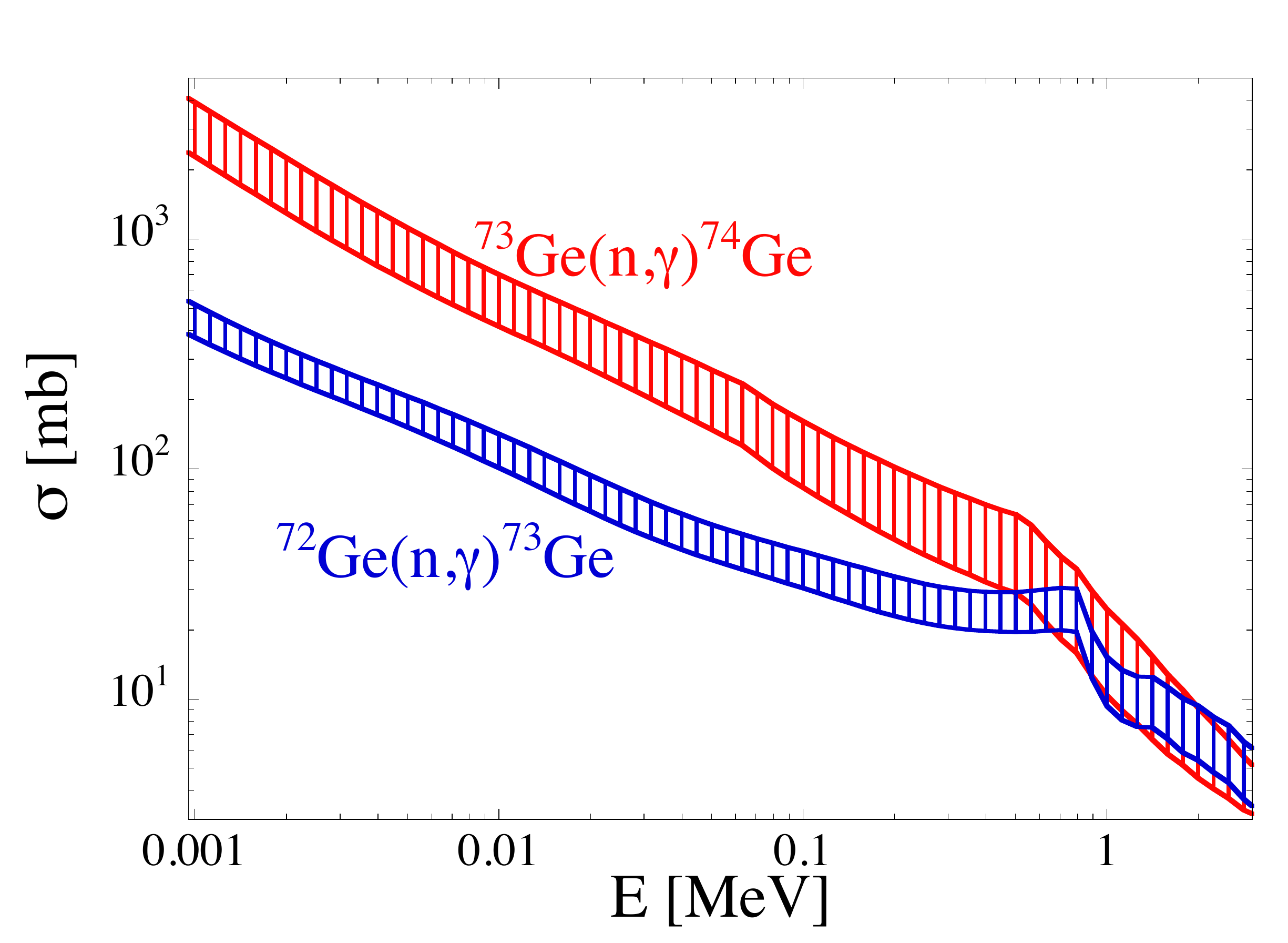}

\caption{(Color online) Neutron capture cross section, $\sigma$, as a function of neutron energy, E, for the $^{72}$Ge(n, $\gamma$)$^{73}$Ge and $^{73}$Ge(n, $\gamma$)$^{74}$Ge reactions. The calculations are performed using TALYS. See text for details.}
\label{fig:crosssec}
\end{center}
\end{figure}

It is interesting to notice that both~\cite{Kap_s_process} and~\cite{Bao_MACS} list $^{72,73}$Ge as amongst the very few of the 277 stable isotopes that at present, lack (n,$\gamma$) cross-section data. Based on our new data on $^{73,74}$Ge, we can provide a semi-experimental capture reaction cross section. The reaction code TALYS-1.6~\cite{talys} is used to perform the calculations. In the case of the neutron-nucleus optical potential we use the Koning $\&$ Delaroche model~\cite{Koning_Potensial}. Based on the present status of the discussion of the electromagnetic character of the upbend, e.g. whether it is of $E$1 or $M$1 type, we treat the input $\gamma$SF in two ways:

\begin{itemize}
\item[(i)] The upbend of the $\gamma$SF is described by the exponential function
\begin{equation} 
f_{up}(E_{\gamma}) = C\rm{exp}\left[-\eta E_{\gamma}\right].
\label{eq:shape_upbend}
\end{equation}
The low-energy enhancement is considered to be of $M1$ type, supported by recent publications~\cite{RSchwengner, ABrown} and the present shell-model calculations. This is combined with QRPA calculations of the $E$1 strength from~\cite{SGoriely_QRPA} and a standard treatment of the $M1$ spin-flip resonance as described in the  TALYS documentation~\cite{talys}. This combined function represents the $\gamma$SF input. 
\item[(ii)] We give measured experimental points of the $\gamma$SF as input, and assume that all the strength is of $E$1 type,~in accordance with~\cite{Litvinova}. 
\end{itemize}
The $\gamma$SF of (i) and (ii) are both combined with two prescriptions for NLD; the one resulting from \emph{norm-1} and the pure HFB + combinatorial one from~\cite{Goriely_level2008}, the uncertainties of the $D_0$ values and $\left<\Gamma_{\gamma}\right>$ values being taken into account. 
In Fig.~\ref{fig:crosssec}, the $^{72}$Ge(n,$\gamma$)$^{73}$Ge and $^{73}$Ge(n,$\gamma$)$^{74}$Ge radiative neutron capture cross sections obtained with the different combinations of NLD and $\gamma$SF are shown. The band width represents the obtained uncertainties taking into account the different input combinations and all systematic and statistical errors.



The corresponding astrophysical Maxwellian-averaged cross sections (MACS) amount to $\left< \sigma \right>$ = 66(13)~mb and 294(78)~mb for $^{72}$Ge(n,$\gamma$) and $^{73}$Ge(n, $\gamma$), respectively, at $k_BT$ = 30 keV (i.e. a temperature of $T = 3.5 \times10^8$ K), and agree well with the previous theoretical values recommended by Bao {\it et al.}~\cite{Bao_MACS} at $\left< \sigma \right>$ = 73(7)~mb and 243(47)~mb, respectively. 

\subsection{Reaction rates of neutron rich Ge isotopes}
It has previously been shown that the upbend can have a significant effect on the neutron capture cross section of exotic neutron-rich nuclei~\cite{AC&Goriely}. Naturally, it is an open question whether the upbend exists in neutron-rich Ge isotopes (Ref.~\cite{AC_76Ge} reports a similar strength of the upbend in the $\gamma$SF of the slightly more neutron rich isotope $^{76}$Ge). The shell-model calculations of the $^{80}$Ge $M1$ $\gamma$SF supports the assumption of a persisting upbend for neutron-rich Ge isotopes. In the following we will assume that the upbend remains as strong in isotopes approaching the neutron drip line as observed in $^{73,74}$Ge. A fit to the $^{74}$Ge $\gamma$SF data give the parameters $(C, \eta)$ = $(4\times 10^{-8}, -0.99)$ in Eq.~(\ref{eq:shape_upbend}). These parameters for the upbend have been applied to the neutron-rich Ge-isotopes. We calculate the ratio of the reaction rates including and excluding the upbend for temperatures corresponding to two proposed r-process sites~\cite{Goriely_rprocess}; a cold r-process in neutron star mergers and a hot r-process as in a neutrino-driven wind in core-collapse. As one goes to neutron rich nuclei we rely on theoretical calculations of the $S_n$, NLD, and $\gamma$SF, since they are at this point experimentally inaccessible. Some uncertainties arise, especially from the mass model used to establish $S_n$. The input used in the TALYS calculations of the astrophysical reaction rates in this case are: for the mass, the Skyrme-HFB mass model of Ref.~\cite{SGoriely_mass}, for the NLD, the HFB + Combinatorial model~\cite{Goriely_level2008} and for the $E1$ $\gamma$SF, the HFB + QRPA model~\cite{SGoriely_E1}. For the $M1$ spin-flip resonance the standard TALYS treatment has been applied~\cite{RIPL3}. 
\begin{figure}
\begin{center}
\includegraphics[clip,height=10cm]{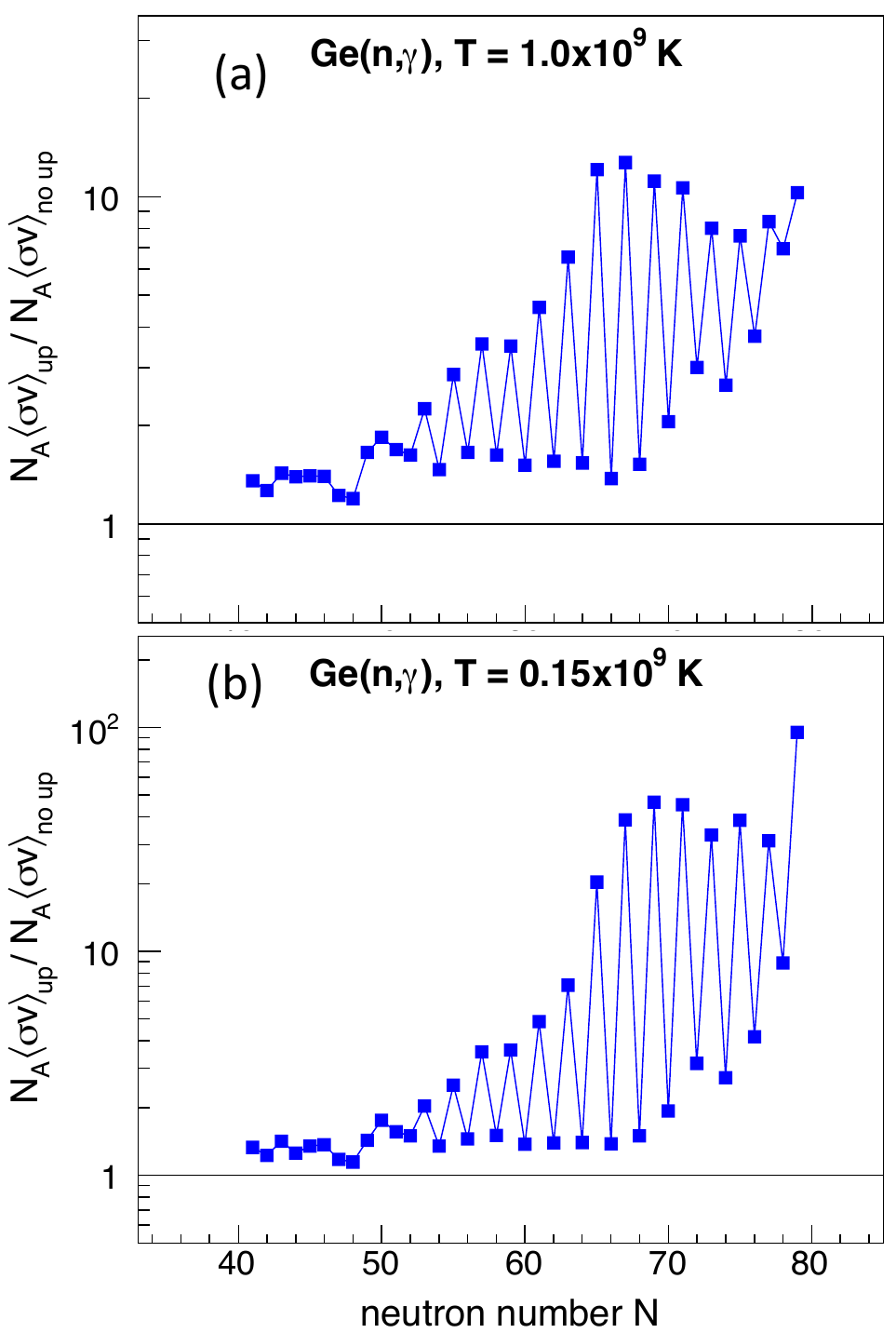}
\caption{(Color online) Ratio of Maxwellian-averaged (n,$\gamma$) reaction rates including and excluding the upbend for the Ge isotopic chain up to the neutron drip line for temperatures (a) $T =  1$ GK and (b) $T =  0.15$ GK.}
\label{fig:reaction_astro}
\end{center}
\end{figure}
Calculations for a hot astrophysical environment of $T = 1.0$ GK are shown in Fig.~\ref{fig:reaction_astro}(a). Odd-even staggering effects are strong, reflecting the difference in the neutron separation energy of isotopes with odd and even neutron number. The sensitive energy range of the $\gamma$SF in the neutron capture reaction is situated typically a few MeV below $S_n$, corresponding to the upbend-region for the extremely neutron-rich odd Ge isotopes. As expected, we see that the influence of the upbend becomes more important as the number of neutron increases. The maximum increase for the extremely neutron rich nuclei is $\approx$ a factor 15 for the case of $T = 1.0$ GK. Figure~\ref{fig:reaction_astro}(b) shows the calculated reaction rates for the case of a cold r-process with $T = 0.15$ GK. For this temperature, an increase of a factor $\approx$ 60-70 in the reaction rates is seen for the most neutron-rich isotopes. Even for more moderately neutron-rich nuclei, an increase of a factor of 2 can be observed.
Hence, we conclude that the impact of the upbend on the (n, $\gamma$) reaction rates could be significant for the Ge case, as already shown in Ref.~\cite{AC&Goriely}.

\section{Summary}
\label{sec:conc}
The NLDs and $\gamma$SFs of $^{73,74}$Ge in the energy range below $S_n$ have been extracted from particle-$\gamma$ coincidence data using the Oslo method. Moreover, the $\gamma$SF above $S_n$ of $^{74}$Ge has been deduced from a photo-neutron experiment. A low-energy enhancement in the $\gamma$SF is observed in both nuclei. Shell-model calculations on $^{74}$Ge indicate that the enhancement is (at least partly) due to $M1$ transitions. The neutron capture cross sections $^{72}$Ge(n,$\gamma$) and $^{73}$Ge(n,$\gamma$) could for the first time be experimentally constrained using our new data as input. The effect of the upbend on the astrophysical reaction rates is investigated, and is shown to be significant for neutron-rich isotopes.  

\vskip 1.5 cm
\noindent{\bf Acknowledgements} \\
The authors wish to thank J.~C.~M\"{u}ller, E~.~A.~Olsen, A.~Semchenkov, and J.~Wikne at the Oslo Cyclotron Laboratory for providing excellent experimental conditions and LBNL for providing the $^{74}$Ge target used in the OCL experiment. A.~C.~L. acknowledges financial support from the Research Council of Norway, project grant no.~205528 and from the ERC-STG-2014 under grant agreement no.~637686. Part of this work was supported by the National Research Foundation of South Africa under contract nos.: 92789 and 78766. D.~M.~F., O.~T. and I.~G. acknowledge financial support from the Extreme Light Infrastructure Nuclear Physics (ELI-NP) Phase I, a project co-financed by the Romanian Government and the European Union through the European Regional Development Fund (425/12.12.2012, POS CCE, ID 1334 SMIS-CSNR 40741). S.~G. acknowledges the financial support from the F.~N.~R.~S.

\end{document}